\documentclass[11pt,twoside]{article}
\usepackage{asp2006}
\usepackage{epsf}
\usepackage{graphics}
\usepackage{lscape}
\def\edcomment#1{\iffalse\marginpar{\raggedright\sl#1\/}\else\relax\fi}
\reversemarginpar

\begin{document}
\title{Three Dimensional Line Transfer Study of Energy Sources in Compact
Molecular Gas in Active Galaxies \\ --- AGN/Starburst Connection ---}
\author{Masako Yamada \altaffilmark{1}}%\footnote{present address : Institute of Astronomy and Astrophysics, Academia Sinica, P.O. Box 23-141, Taipei 10617
%, Taiwan, R.O.C.} 
\affil{ALMA-J Office, National Astronomical Observatory of Japan, Mitaka, Tokyo, 181-8588, Japan}
\altaffiltext{1}{present address : Institute of Astronomy and Astrophysics, Academia Sinica, P.O. Box 23-141, Taipei 10617, Taiwan, R.O.C.}
\author{K. Wada and K. Tomisaka}
\affil{Center for Computational Astronomy, National Astronomical 
Observatory of Japan, Mitaka, Tokyo, 181-8588, Japan}

\begin{abstract}
Recent observations suggest molecular line ratios in millimeter and submillimeter 
bands may be a good tool to reveal the long-standing question on the origin of 
energy sources in obscured active galaxies -- AGN and/or starburst.
%On the other hand, 
Observations of actual molecular medium show in general inhomogeneous 
structures as well as high-resolution hydrodynamic simulations do.
In order for precise interpretation of emergent line emission from the inhomogeneous
molecular gas to probe the dominant energy source of active galaxies, we 
study characteristic features of emergent intensities via three-dimensional non-LTE (non-local
thermodynamic equilibrium) line transfer simulations.
Our results succeeded in making clear 1)  the necessary conditions for HCN/HCO$^{+}$-dichotomy,
and 2) importance of clumpiness on intensity ratio and its interpretation. 
These results are obtained for the first time by our realistic three-dimensional simulations,
and line transfer simulations will be a powerful tool to comprehensive studies of 
extragalactic interstellar medium (ISM) in forthcoming ALMA (Atacama Large 
Millimeter/submillimeter Array) era.
\end{abstract}

\keywords{active galactic nuclei -- starburst -- ISM -- radiative transfer}

\vspace{-0.5cm}
\section{Introduction}

Molecular gas in external galaxies is a subject of crucial importance for observational 
and theoretical studies of galaxy formation.
Among the interstellar medium (ISM), compact molecular gas around an active 
galactic nuclei (AGN) is expected to be energy budget from AGN and/or the nuclear starburst 
of possible relevance.
Recent observations suggest line ratios in millimeter and submillimeter bands may be 
a good tool to reveal the long-standing question of the origin of activities in obscured 
active galaxies  -- AGN or nuclear starburst \citep[see e.g.][]{kohno2001, kohno2005, imanishi2007}.

Current observations present compact molecular gas of size $\approx$ 1 kpc or less,
though, hydrodynamic simulations predict highly inhomogeneous and clumpy substructures
in a "compact" molecular gas at the centers of active galaxies \citep{wada2001, wada2005}.
In such an inhomogeneous molecular gas, energy level populations of emitting particles
will be also complex and complicated one, which makes derivation of correct 
physical features of emitting ISM from emergent line intensities difficult.
Then we study excitation conditions and line intensities from an inhomogeneous molecular 
gas at the center of an active galaxy, with three-dimensional non-local thermodynamic 
equilibrium (non-LTE) line transfer simulations along with the results of high-resolution 
hydrodynamic simulation \citep{wada2005, yamada2007b, yamada2008b}.

%%%
\section{Numerical Calculations}

We first performed hydrodynamic simulations of a compact ($R\le 32$ pc) ISM in a 
steady gravitational potential of supermassive blackhole ($M_\mathrm{SMBH}=10^8M_\odot$)
and host galaxy. 
And then we performed non-LTE line transfer simulations as a post process using 
a snapshot result of hydrodynamic simulation.
We solved non-LTE rate equations in statistical equilibrium up to $J=10$ and 
integrated a standard radiative transfer equation along all the sampling rays to 
calculate mean intensity $\bar{J}\equiv 1/(4\pi)\int I_\nu d\Omega$.
We calculated these two equations iteratively until energy level distributions and 
mean intensity field converge with $\approx 10^{-6}$ relative precision (\citealt{hogerheijde2000}
; see \citealt{yamada2007b} for details).

%%%%
\section{HCN and HCO$^{+}$ Dichotomy of Active Galaxies}

Since the pioneering work of \citet{kohno2001}, it has been argued that intensity 
ratio of HCN and HCO$^{+}$ rotational lines would be able to probe obscured 
central energy sources of active galaxies in terms of chemical abundances 
\citep[see e.g.][]{kohno2005, imanishi2007, baan2008}.
The basic idea underlying this dichotomy is that strong X-ray from AGN will form 
X-ray dominated regions (XDR) compared with stellar feedback from nuclear 
starburst that will form photodissociation region (PDR), and different chemical 
structures in XDR and PDR imprinted in molecular line ratio indirectly enables 
unveil the central energy sources.

\begin{figure}[tbhp]
 \plotone{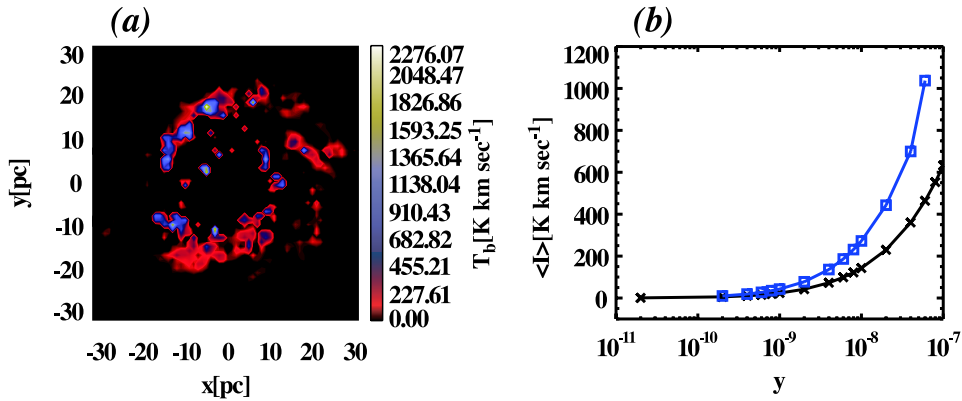}
  \caption{Left panel : integrated intensity of HCN (1-0) line in a face-on view. right 
  panel : average integrated intensity $\langle I \rangle$ for HCN (black line with crosses)
  and HCO$^{+}$ (blue line with squares). }
  \label{fig:result1}
\end{figure}
In Figure \ref{fig:result1} we show our numerical simulation results {\slshape with an 
assumption of spatially uniform chemistry}. 
Distribution of integrated intensity of HCN (1-0) line (Fig. \ref{fig:result1}(a)) becomes 
quite inhomogeneous, reflecting the inhomogeneous structure in the molecular gas 
model. 
Numbers of bright regions of which radius is $\approx$ 10pc or less are observed.
Current millimeter observational instruments cannot resolve these small structures except
for the closest galaxies, but forthcoming ALMA (Atacama Large Millimeter/submillimeter
Array) telescope will reveal these internal structures in a "compact" gas at the center of 
a distant galaxy. 
In order to examine intensity ratio of HCN(1-0) and HCO$^{+}$ (1-0), we computed 
HCN and HCO$^{+}$ lines for a wide range of molecular abundances ($10^{-11}
\le y \le 10^{-7}$, where $y$ is relative molecular abundance to H$_2$).
In Figure \ref{fig:result1} (b) we plot the ratio of resultant intensities averaged over 
the "field-of-view" of the simulation as a function of $y$ for face-on view. 
Our results show that for the same value of $y$, HCO$^{+}$ (1-0) is always brighter than 
HCN (1-0), so that a high intensity ratio $R_{\mathrm{HCN/HCO}{+}}$ greater than unity
requires overabundance of HCN compared with HCO$^{+}$ ($y_\mathrm{HCN} 
\sim 10\times y_{\mathrm{HCO}^{+}} $) on average.

Current models of XDR chemistry predicted {\slshape under-abundance} of HCN
\citep{meijerink2005, meijerink2007}, and then original picture of HCN/HCO$^{+}$ 
diagnostics in terms of XDR/PDR chemistry would need revision and sophistication.
Extended observations have given rise discussion on alternative mechanisms that 
determine $R_{\mathrm{HCN/HCO}{+}}$ such as mid-IR photon pumping \citep{aalto2007}.
Extended line transfer simulations including non-uniform chemical structure and/or
continuum pumping will serve as a powerful tool to more detailed studies as well as
more sophisticated chemistry models with realistic gas geometry.

%%%%
\section{Intensity Ratio of Millimeter and Submillimeter Lines}

Recent progress of submillimeter observations of high-$J$ molecular lines opened 
a way to examine physical conditions of emitting ISM in more detail.
Since critical densities for LTE population and wavelengths are dependent on $J$, 
average intensity ratio will be affected by density and temperature structures, and 
the effect of inhomogeneity is expected to be stronger for high density tracer 
because of small fraction of dense and thermalized gas.
We examined the ratio of $J=$4-3 and 1-0 lines of HCN molecule as representative 
dense gas tracers with line transfer simulations described in \S2 \citep{yamada2008a, 
yamada2008b}.

\begin{figure}[tbhp]
\plotone{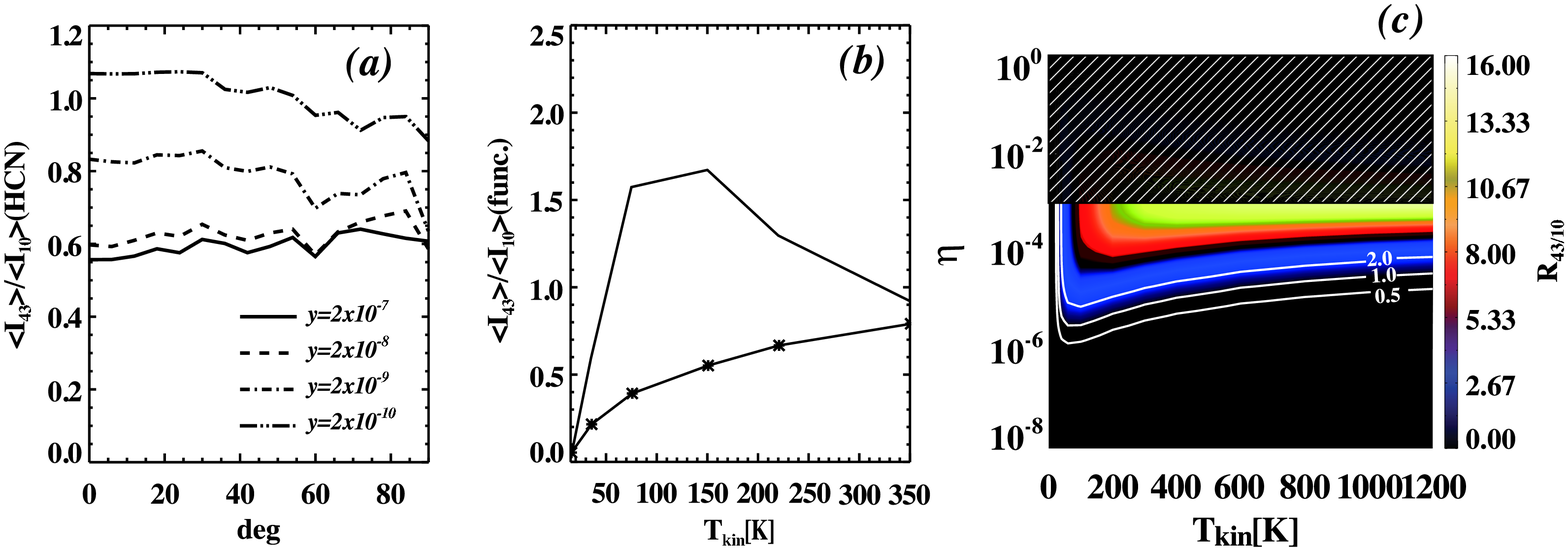}
  \caption{Panel (a) : average intensity ratios $R_{43/10}$ of HCN for four abundance values
  taken from line transfer simulations. Panel (b) : comparison of analytic formula (solid line) for
  a simple isothermal two-phase model and numerical line transfer results of isothermal ISM models.
  In panel (c) numerically evaluated $R_{43/10}$ for an isothermal two-phase modelling are plotted
  as a function of $\eta$ and $T_\mathrm{kin}$, and a fixed average density $n_\mathrm{ave}=100$
  cm$^{-3}$. }
  \label{fig:result2}
\end{figure}
Figure \ref{fig:result2} (a) shows averaged intensity ratio as a function of viewing angle
$\theta$ ($\theta=0^{\circ}$ is face-on view, and 90$^{\circ}$ is edge-on view).
It shows that average $R_{43/10}$ decreases with increase of $y$, which contradicts 
one-zone estimation that $R_{43/10}$ approaches unity as $y$ (or optical thickness) increases.
This result is likely to come from multi-phase nature of emitting ISM.

%--
\subsection{Two-Phase Modelling and Analytic Formula of $R_{43/10}$}

We model multi-phase ISM as a simple two component one 
based on the knowledge of excitation mechanisms.
Since we know that average kinetic temperature of model ISM (174 K) is sufficiently higher 
than the excitation energy of $J=4$ level of HCN, we assume single temperature ISM.
Furthermore in order for analytic examination we assume optically thin over a whole region.
Then our two-phase model consists of two density components, one is dense 
($n\ge n_\mathrm{crit}$) clumps of which volume fraction is $\eta$,
 and another is tenuous ($n\ll n_\mathrm{crit}$) ambient.

Considering that in dense ($n\ge n_\mathrm{crit}$) regions level population is close to
LTE, and in warm ($T_\mathrm{kin} \gg h\nu_{J,J-1}/k_B$) and tenuous ($n\ll n_\mathrm{crit}$)
regions level population is a balance of spontaneous decay and collisional excitation \citep{yamada2007a}, average intensity ratio $R_{43/10}$ is described as a 
ratio of sum of intensities from both regions,
\begin{eqnarray}
  R_{43/10} &\simeq& \displaystyle
    \frac{\sum_\mathrm{vol}(n_{\mathrm{H}_2}yf_4A_{43}h\nu_{43})\nu_{43}^{-2}}{\sum_\mathrm{vol}(n_{\mathrm{H}_2}yf_1A_{10}h\nu_{10})\nu_{10}^{-2}}
    = \left(\frac{\nu_{43}}{\nu_{10}}\right)^{-1} \times 
        \frac{\sum_\mathrm{vol}(n_{\mathrm{H}_2}f_4A_{43})}{\sum_\mathrm{vol}(n_{\mathrm{H}_2}f_1A_{10})},  \label{eq:2phaseR} \\
    &\propto& 
    \frac{\eta_{43}(n_{\mathrm{H}_2}f_4A_{43})_d +
  (1-\eta_{43})(n_{\mathrm{H}_2}f_4A_{43})_t}{\eta_{10}(n_{\mathrm{H}_2}f_1A_{10})_d
  +(1-\eta_{10})(n_{\mathrm{H}_2}f_1A_{10})_t}.  \label{eq:defA}
\end{eqnarray}
Figure \ref{fig:result2} (c) shows an numerical evaluation of thus obtained $R_{43/10}$.
It is obvious that average $R_{43/10}$ is quite sensitive to $\eta$, and {\slshape can take almost 
any value up to theoretical maximum (=16 in this case) 
even if average density ($n_\mathrm{ave}=100$ cm$^{-3}$) is smaller than critical density 
($n_\mathrm{crit} = 10^5$ cm$^{-3}$ for $J=$1-0 line)}.
In other words, intensity ratio close to unity does not guarantee a thermalized population.

In Figure \ref{fig:result2} (b) we plot numerical evaluation of equation (\ref{eq:2phaseR}) with 
a definition of $\eta_{43}=\eta_{10}=\eta_\mathrm{sim}$
\begin{equation}
  \eta_\mathrm{sim} 
  \equiv 
  \frac{N_\mathrm{grids}(n_1/n_o \ge \epsilon
  g_J/g_0\exp{(-E_{10}/k_BT_\mathrm{kin})})}{N_\mathrm{grids}}. 
  \label{eq:etasim}
\end{equation}
Compared with results of line transfer simulations of isothermal hydrodynamic data 
generated by original input model, we can see that the analytic formula of $R_{43/10}$ 
(\ref{eq:2phaseR}) can fit 
numerical simulation results within a factor of two in spite of simplicity of this formula.
We thus concluded that multi-phase modelling based on excitation condition considerations
along with numerical line transfer simulations could help construct more sophisticated ways to 
derive correct physical features and structures from observed line intensities.
Our results also demonstrated the importance of spatial structure of density (and temperature)
on emergent intensity besides chemical structure.

%%%
\section{Summary}

We showed theoretical considerations on molecular line diagnostics of active galaxies
by numerical experiments of three-dimensional line transfer simulations.
Two kinds of results ($R_{\mathrm{HCN/HCO}^{+}}$ and $R_{43/10}$ of HCN) showed that
simple models need revisions to take into account of inhomogeneities of emitting ISM.
Numerical experiments can achieve a high-resolution examination in prior to the 
forthcoming ALMA project, and will be a powerful tool to examine physics imprinted in
emergent intensities.

%%%%
\acknowledgements
Numerical computations were partly carried out on the PC cluster at Center for 
Computational Astrophysics (CfCA), National Astronomical Observatory of Japan, 
and this research was supported in part by Grants-in-Aid by the Ministry of Education,
Science, and Culture of Japan (16204012).

%%%%

\end{document}